# Cosmologia dionisíaca*

Juliano Neves**

**Resumo:** Destaca-se na filosofia nietzschiana a importância do conceito de força proveniente da física para a construção de um dos seus principais conceitos: a vontade de potência. O conceito de força, que Nietzsche buscou na mecânica clássica, quase desaparece na física do século XX com a teoria quântica de campos e a teoria da relatividade geral. Ainda é possível, na ciência de hoje, o *mundo* nietzschiano como forças em disputa, uma cosmologia dionisíaca?
**Palavras-chave:** ciência - cosmologia – vontade de potência – forças

## 1. Vontade de potência

É, entre outras leituras científicas, no multifacetado Roger Boscovich (físico, matemático e astrônomo croata) que Nietzsche depara com o conceito de força na física. De acordo com Jammer, físico e historiador da ciência, Boscovich estudou fenômenos de colisão dos corpos no século XVIII e teria concluído que "impenetrabilidade e extensão (…) são meramente expressões espaciais de forças, 'força' é consequentemente mais fundamental do que matéria (…)."[1] Em *Além do bem e do mal*, Nietzsche escreve: "(...) Boscovich nos ensinou a abjurar a crença na última parte da terra que permanece firme, a crença na 'substância', na 'matéria', nesse resíduo e partícula da terra, o átomo (…)" (JGB/BM 12, KSA 5.26. Trad. PCS)[2]. No entanto, o conceito de força da física era, para o filósofo alemão, vazio. Num fragmento póstumo, afirma: "Uma força que não podemos conceber (como a assim chamada força de atração e de repulsão puramente mecânica) é uma palavra vazia (…). Todo acontecer derivado de propósitos é redutível ao *propósito de ampliar a potência*" (Nachlass/FP 1885-86, 2[88], KSA 12.105. Trad. FRK)[3]. Então, no conceito de força insere um mundo interior: a vontade de potência. "O vitorioso conceito de 'força', com o qual nossos físicos recriaram Deus e o mundo, necessita ainda de uma complementação: é necessário aditar-lhe um mundo interior, ao qual eu chamo de 'vontade de potência'" (Nachlass/FP 1885, 36[31], KSA 11.563. Trad. FRK).

Muito já foi escrito sobre o conceito de vontade de potência em Nietzsche. Müller-Lauter[4], por exemplo, discute uma possível ou não interpretação metafísica desse importante conceito na filosofia nietzschiana. Marton[5], por outro lado, destaca a influência não apenas de Boscovich mas também de outros cientistas, como os biólogos

---

[1] Cf. JAMMER, M. *Concepts of force*. New York: Dover Publications Inc., 1999, p. 178 (as traduções feitas do inglês são de minha responsabilidade). Para uma discussão da influência de Boscovich sobre Nietzsche, cf. ITAPARICA, A. L. M. Nietzsche e Boscovich: dinamismo e vontade potência. In: Azeredo, V. D. *Encontros Nietzsche*. Ijuí: Editora Unijuí, 2003, p. 163-171.

[2] Indicamos os tradutores da obra de Nietzsche em português pelas iniciais dos seus nomes e sobrenomes nas passagens citadas: PCS (Paulo César de Souza), RRTF (Rubens Rodrigues Torres Filho), FRK (Flávio R. Kothe) e PS (Pedro Süssekind).

[3] Nas traduções de FRK, substituímos a expressão "vontade de poder" por "vontade de potência". Além disso, preferimos força (para a tradução de *Kraft*) ao invés de energia. Isso se deve porque consideramos que com o uso do conceito de energia perde-se a noção de luta e disputa, pelo menos enquanto a cosmologia einsteiniana não entra em jogo, como veremos a seguir.

[4] Cf. MÜLLER-LAUTER, W. *A doutrina da vontade de poder em Nietzsche*. Trad. Oswaldo Giacoia Junior. São Paulo: Annablume, 1997, p. 70.

[5] Cf. MARTON, S. *Nietzsche: das forças cósmicas aos valores humanos*. 3. ed. Belo Horizonte: Editora UFMG, 2010, cap. 1.

Rolph e Roux, na criação do conceito. Se em *Assim falou Zaratustra*[6] a vontade de potência refere-se ao mundo orgânico, em sua obra posterior refere-se também ao mundo inorgânico. É o filósofo que diz como vê o mundo, um mundo como forças e vontade de potência: "E sabeis sequer o que é para mim 'o mundo'? (...) Este mundo: uma monstruosidade de força, sem início, sem fim, uma firme, brônzea grandeza de força, que não se torna maior, nem menor, (...) jogo de forças e ondas de forças ao mesmo tempo um e múltiplo (...) – *Esse mundo é a vontade de potência – e nada além* disso!" (Nachlass/FP 1885, 38[12], KSA 11.610. Trad. RRTF). Um mundo de forças que combatem entre si por mais potência. A cosmologia nietzschiana é tida como uma cosmologia das forças e da disputa. Forças que não aumentam e não diminuem na totalidade (Nietzsche simpatizava com a lei da conservação de energia porque pensava na conservação das forças, que exigia, promovia ou estimulava, segundo o filósofo, a tese do eterno retorno[7]). E numa eternidade imanente, tais forças repetem os seus arranjos – nasce, a partir daí, a ideia do eterno retorno do mesmo em sua versão cosmológica.

Neste trabalho, o tema não é o eterno retorno do mesmo[8]. Queremos aqui, como já fizemos, destacar a importância do conceito de força proveniente da física para construção do *mundo* nietzschiano e, em seguida, mostrar que, mesmo sem tal conceito, a cosmologia hoje ainda pode fortalecer a visão de mundo do filósofo alemão. E os ingredientes dessa visão, além do conceito de força, são: o *ágon* (a disputa) e a impermanência, que são subsumidos no conceito de vontade de potência.

---

[6] "Apenas onde há vida há também vontade: mas não vontade de vida, e sim – eis o que te ensino – vontade de potência!" (Za/ZA II, *Da superação de si* mesmo, KSA 4.149. Trad. PCS).
[7] "A tese da permanência da energia estimula o *eterno retorno*" (Nachlass/FP 1886, 5[54], KSA 12.205. Trad. FRK).
[8] Cf. NEVES, J. O eterno retorno hoje. In: *Cadernos Nietzsche*, vol. 32, p. 283-296, 2013, para uma discussão do eterno retorno nietzschiano do ponto de vista cosmológico e a sua proximidade com cosmologias científicas atuais além do modelo padrão.

## 2. O ocaso do conceito de força na física

A partir da física do século XX, o conceito de força é praticamente deixado de lado. Na teoria da relatividade geral de Einstein, sua teoria da gravitação, o cálculo da órbita de um corpo de teste[9] ao redor de uma estrela, por exemplo, dispensa o conceito de força. Bastam a métrica do espaço-tempo que descreve tal estrela e as equações da geodésica, que podem ser obtidas de considerações puramente geométricas. Como resultado, temos as trajetórias possíveis do corpo de teste em órbita ao redor da estrela submetido somente à gravitação. Gravidade em Einstein pode ser vista não como uma força, mas como uma "deformação" no espaço-tempo. Mesmo no mundo quântico, o modelo padrão que descreve as partículas subatômicas utiliza o conceito de interação ao invés do de força. As quatro "forças" ou interações fundamentais na física de hoje – eletromagnética, nuclear forte, nuclear fraca e gravitacional[10] – são descritas por partículas chamadas bósons de calibre. Por exemplo: a chamada força elétrica no modelo padrão das partículas é descrita pela interação, pela troca, de fótons, que são os bósons de calibre da interação eletromagnética. Segundo Jammer, "(...) o que chamamos de as 'quatro forças fundamentais da natureza' não são 'forças' no sentido tradicional. Em suma, a física de partículas moderna, assim como a teoria da relatividade geral, parece apoiar a tese de que o conceito de força alcançou o fim do seu ciclo vital (...)"[11]. Com a *perda de força do conceito de força*, realiza-se um trajeto completo do nascimento ao ocaso de tão importante noção. "Com os trabalhos de Mach, Kirchhoff e Hertz, o processo de eliminação do conceito de força da mecânica completou o seu desenvolvimento lógico."[12] Processo que se torna bem-sucedido no século XX.

---

9  Por corpo de teste queremos dizer que sua massa é muito menor do que a massa da estrela envolvida.
10  Esta última ainda não está contida no chamado modelo padrão das partículas pois o campo gravitacional é o único ainda não quantizado satisfatoriamente. Para isso, é necessária a tão buscada e desejada teoria quântica da gravitação.
11  Cf. JAMMER, M. *Concepts of force*. New York: Dover Publications Inc., 1999, p. vi.

Nesse sentido, cabe a pergunta: a filosofia nietzschiana perde algo com o ocaso do conceito de força na física? Ao nosso ver, num patamar acima do conceito de força estão os conceitos de *ágon* e de mudança na filosofia nietzschiana. Num texto de juventude, o filósofo escreve sobre a importância do ágon, da disputa, na cultura grega antiga: "Para os antigos (…) o objetivo da educação 'agônica' era o bem do todo, da sociedade citadina" (CV/CP, *A disputa de Homero*, KSA 1.789. Trad. PS). Em Heráclito, Nietzsche elogia a mudança, a impermanência, o vir-a-ser, a disputa ou guerra e reconhece um parentesco com o filósofo antigo, descrevendo o seu mundo dionisíaco: "A afirmação do fluir *e do destruir*, o decisivo numa filosofia dionisíaca, o dizer Sim à oposição e à guerra, o *vir a ser*, com radical rejeição até mesmo da noção de 'Ser' – nisto devo reconhecer, em toda circunstância, o que me é mais aparentado entre o que até agora foi pensado." (EH/EH, *O nascimento da tragédia* 3, KSA 6.313. Trad. PCS).

Acreditamos que a importância dada por Nietzsche ao conceito de força oriundo da física deve-se à sua completa imanência e à corroboração de uma visão de mundo agonística (ou agônica) e em eterna mudança. Conceito que Nietzsche usou como uma ferramenta para descrever a sua cosmologia dionisíaca.

Na próxima seção, mostraremos que, mesmo sem o uso do conceito de força, a cosmologia científica atual pode fortalecer a visão de mundo nietzschiana, uma visão dionisíaca, não somente por possibilitar – ainda que de uma forma remota – um eterno retorno de tudo mas por fornecer a imagem da disputa e do fluir num mundo impermanente. Por outro lado, se no macrocosmo, na cosmologia, temos algo que vai na direção da visão dionisíaca acima descrita, no microcosmo, na física de partículas, a noção de substância, *Ser*, algo que permanece, ainda tem vez. São os campos quânticos que fazem esse papel na física. Sendo as partículas manifestações de tais campos quânticos.

---

12  Ibidem, p. 241.

## 3. Eras de dominação

A cosmologia científica hoje é possível em diversos contextos. O principal deles é o da teoria da relatividade geral de Einstein. Ou seja, em geral, a cosmologia científica, a partir do século XX, parte das chamadas equações de Einstein[13]. Outros ingredientes são usados: o princípio cosmológico, que afirma a igualdade entre observadores num mesmo tempo em diferentes pontos do universo, e a descrição do conteúdo de matéria e energia no universo com o uso de fluidos perfeitos, a partir de uma certa escala de distância, quando o universo se apresenta homogêneo e isotrópico[14]. Atualmente, o mais bem-sucedido modelo cosmológico na ciência é o Lambda-CDM, também conhecido como modelo padrão cosmológico, que tem como ingredientes extras: a matéria escura fria (Cold Dark Matter), um tipo de matéria que só interage gravitacionalmente; a constante cosmológica[15], Lambda, que nesta cosmologia representa a energia escura (a fonte da atual expansão acelerada do cosmo, detectada em 1998); por fim, a chamada fase

---

13  Outras, como a cosmologia de Branas, usam equações do campo gravitacional que são ainda mais gerais do que as equações de Einstein. No entanto, são cosmologias mais ousadas, que chegam a utilizar dimensões extras, por exemplo.
14  Como sugerimos em *O eterno retorno hoje*, o princípio cosmológico é um eco de um ideal moderno, a democracia. O uso de fluidos perfeitos em cosmologia mostra-se mais simples e, acima de tudo, funcional. Um universo isotrópico significa: a física é a mesma em qualquer direção.
15  Esta é a mesma constante que Einstein adicionou às suas equações para obter um universo estático. Com a observação feita por Hubble em 1929 da expansão cósmica e com a demonstração de que o universo estático einsteiniano é instável, o físico alemão abandonou sua constante. Décadas depois, no fim do século XX, ressurge como uma possível fonte da expansão cósmica que, desde de 1998, é tida como acelerada.

inflacionária, que teve início e fim nos primórdios da atual fase de expansão do universo[16].

Apesar dos seus inúmeros feitos, e os recentes dados do telescópio espacial Planck os corroboram, o modelo padrão cosmológico apresenta problemas. Dentre estes, podemos apontar talvez aquele que seja o principal: o problema da singularidade inicial. Este é um problema comum da teoria de Einstein. Seja no contexto cosmológico ou no interior de buracos negros, singularidades surgem como limitações da física einsteiniana. Na física de Einstein, uma singularidade significa uma incapacidade de a física prever um valor finito para uma grandeza. Na cosmologia, a singularidade inicial chama-se *big bang*. De acordo com o modelo padrão cosmológico, há quase 14 bilhões de anos, o universo surge a partir de um evento singular. Nele, o *big bang*, grandezas físicas como a densidade de energia divergem – assumem um valor ilimitado. Espera-se que a solução definitiva para o problema das singularidades na física seja alcançada pela ainda não construída teoria quântica da gravitação. Enquanto isso, sem a desejada teoria, para lidar com tal problema desafiador, surgiram as cosmologias com ricochete (*bouncing cosmologies*[17]). Nestas, a singularidade inicial é evitada com violações de condições de energia impostas pelos chamados teoremas de Hawking-Penrose. Tais teoremas afirmam que singularidades são inevitáveis na teoria einsteiniana, desde que a matéria e a energia do universo obedeçam a determinadas condições. No entanto, a chamada energia escura pode violar condições desse tipo. Dessa forma, a partir de sua observação em 1998, ricochetes são mais plausíveis dentro da física. E ricochete, neste contexto, significa a transição não singular – sem o problema da singularidade inicial – de uma fase de contração para uma fase de expansão (a atual fase cósmica). Numa visão cíclica, o universo expande-se e contrai-se eternamente, passando por ricochetes. Sendo assim, como já tratamos em *O eterno retorno hoje*, um eterno retorno do mesmo, como Nietzsche pensou, torna-se um assunto digno de discussão na cosmologia científica.

---

16 A inflação ou fase inflacionária foi um período de expansão rapidíssima, onde características hoje observadas, como a isotropia e a homogeneidade cósmicas, foram geradas. Além disso, a inflação é um bem-sucedido mecanismo de geração de estruturas. Ou seja, as sementes que geraram estruturas como galáxias surgiram na fase inflacionária.

17 Cf. NOVELLO, M. & BERGLIAFFA, S. E. P. "Bouncing cosmologies". In: *Physics reports*, vol. 463, p. 127-213, 2008, para uma revisão técnica e cf. NOVELLO, M. *Do big bang ao universo eterno*, 2. ed. Rio de Janeiro: Zahar, 2010, para uma introdução não técnica ao tema das cosmologias sem a singularidade inicial. Igualmente a física de buracos negros sofre do mesmo problema. Recentemente, Saa e eu construímos métricas que descrevem buracos negros não singulares (regulares), generalizando e ampliando resultados preexistentes (cf. NEVES, J. C. S. & SAA, A. "Regular rotating black holes and the weak energy condition". In: *Physics letters B*, vol. 734, p. 44-48, 2014).

Na cosmologia padrão ou em cosmologias com ricochete, a dinâmica do tecido do espaço-tempo é governada pela equação de Friedmann. Essa equação é obtida a partir das equações do campo gravitacional, ou equações de Einstein, com o uso de um ingrediente essencial: a métrica de Friedmann-Lemaître-Robertson-Walker (é a métrica que descreve se um espaço-tempo é finito, infinito, limitado ou ilimitado). A equação de Friedmann descreve a expansão (que pode ser acelerada ou não), uma possível contração ou até mesmo a estaticidade do tecido do espaço-tempo. Tal equação possui alguns termos: há aquele que representa a matéria (escura e ordinária, sendo esta última aquela de que estrelas, planetas e a vida são compostos), outro que representa a radiação e até o termo que representa a energia escura. Em cada período do universo, um termo é o mais importante ou relevante, ou seja, tem mais peso na equação de Friedmann. Isto é, no início da atual fase de expansão do universo, o termo da radiação é mais importante do que o termo da matéria ou da energia escura. Nesse sentido, dentro da cosmologia, costuma-se usar a expressão *era de dominação da radiação* para indicar o período inicial da atual fase de expansão cósmica. Atualmente, acredita-se, vivemos a fase em que a energia escura domina. Doravante, será o termo na equação de Friedmann mais relevante, aquele que dominará e conduzirá a expansão cósmica.

Ao nosso ver, a cosmologia científica mostra-nos um interessante quadro: uma descrição do mundo como composto por fluidos. Em cada período da história cósmica há o domínio de um tipo de fluido, aquele que tem mais peso na equação de Friedmann. Numa cosmologia cíclica, as eras de dominação se alternam. Radiação, matéria e energia escura se alternam no domínio. Eternamente, a sequência radiação-matéria-energia escura se repete. Uma interpretação agonística é possível: a disputa entre as formas de matéria e energia não tem término. Este mundo que se cria e se destrói eternamente sem um estado final pode, seguindo Nietzsche, ser chamado de *dionisíaco ou vontade de potência*: "esse meu mundo *dionisíaco* do eternamente-criar-a-si-próprio, do eternamente-destruir-a-si-próprio, esse mundo secreto da dupla volúpia, esse meu 'para além de bem e mal', sem alvo (…) quereis um *nome* para esse mundo? (…) *Esse mundo é a vontade de potência – e nada além disso*!" (Nachlass/FP 1885, 38[12], KSA 11.611. Trad. RRTF).

## 4. Considerações finais

"O vitorioso conceito de 'força'" proveniente da física, utilizado por Nietzsche ora como sinônimo da vontade de potência, ora como algo que necessita de "uma complementação" (e a vontade de potência o preenche como o seu mundo interior), é visto por nós como mais uma de tantas ferramentas usadas pelo filósofo. E mesmo com o seu quase abandono na física, a partir do século XX, com as teorias da relatividade geral e quântica de campos, o *mundo* nietzschiano – como um mundo dionisíaco – é ainda possível na ciência, especialmente na cosmologia. Tendo em vista uma de suas principais equações, aquela que descreve o tecido do espaço-tempo, a equação de Friedmann, sugerimos uma interpretação agonística para as chamadas *eras de dominação*. Sendo assim, um mundo que flui, numa disputa sempiterna entre as formas de matéria e energia, pode ser sugerido pela cosmologia científica atual.

**Abstract:** In the Nietzschean philosophy, the concept of force from physics is important to build one of its main concepts: the will to power. The concept of force, which Nietzsche found out in the Classical Mechanics, almost disappears in the physics of the XX century with the Quantum Field Theory and General Relativity. Is the Nietzschean world as contending forces, a Dionysian cosmology, possible in the current science?
**Keywords:** science – cosmology – will to power – forces

## referências bibliográficas